\documentclass[lettersize,journal]{IEEEtran}
\usepackage{amsmath,amsfonts,mathtools}
\usepackage{algorithmic}
\usepackage{algorithm}
\usepackage{array}
\usepackage[caption=false,font=normalsize,labelfont=sf,textfont=sf]{subfig}
\usepackage{textcomp}
\usepackage{stfloats}
\usepackage{url}
\usepackage{verbatim}
\usepackage{hyperref}
\usepackage{graphicx}
\usepackage{lipsum}
\usepackage{cite}
\usepackage[svgnames]{xcolor}
\usepackage{amssymb}

\begin{document}

\title{How haptic feedback enables human group synchronization}

\author{%
Angelo Di Porzio,~\IEEEmembership{Member,~IEEE,},\thanks{A. Di Porzio is with Scuola Superiore Meridionale, Naples, Italy}
Etienne Burdet,~\IEEEmembership{Member,~IEEE},\thanks{E. Burdet is with Imperial College of Science, Technology and Medicine, SWA 2AZ, London, UK; https://www.imperial.ac.uk/human-robotics/}
Marco Coraggio,~\IEEEmembership{Member,~IEEE}
\thanks{M. Coraggio is with Scuola Superiore Meridionale, Naples, Italy}
}


\maketitle

\begin{abstract}
Synchronization often emerges spontaneously among interacting people, yielding practical benefits for tasks such as sports, physical rehabilitation, and collaborative manufacturing, and fostering a sense of unity and trust. Although
visual interaction is typically considered the primary channel for achieving synchronization, it is constrained by sensorimotor delays, occlusions, and limited attentional resources. Here, we tested whether haptic communication can induce or enhance synchronization in a group motor task. Six quartets performed oscillatory wrist flexion/extension while being connected via virtual elastic bands; we compared haptic only, visual only, combined haptic\&visual, and no feedback conditions. Even weak haptic coupling induced group synchronization, and combining haptic with visual feedback yielded higher and more stable synchronization than either modality alone. Haptic feedback also increased the frequency at which groups coordinated, and movement smoothness rose with frequency. These findings lay the groundwork for designing haptic interaction protocols for collaborative group environments.
\end{abstract}


\section{Introduction}

\IEEEPARstart{S}{ynchronization} naturally emerges between people. It may occur intentionally, when coordinated motion improves task performance, as in team rowing or ensemble music, or spontaneously, as a byproduct of interaction. Visual and auditory feedback have been shown to induce synchronized behavior in tasks such as rocking, clapping, and musical performance \cite{richardson2007rocking, neda2000sound, shahal2020synchronization, shniderman2024how, volpe2016measuring}. Synchronization also carries benefits beyond task performance, fostering unity and trust and promoting well-being, affiliation, and cooperation \cite{rennung2016prosocial, hove2009s, wiltermuth2009synchrony}. Improving synchronization is therefore worth pursuing for both task-related and prosocial reasons.

The emergence of synchronization depends on how effectively people exchange information. Several studies have focused on visual interaction, especially during one-dimensional oscillatory movement tasks. The \emph{multiplayer oscillatory mirror game}, for example, has been widely used to study how people coordinate their motion while observing each other \cite{alderisio2017interaction}. In this paradigmatic task, participants oscillate their hands back and forth along an imaginary straight line while watching each other, trying to synchronize their movements. The simplicity of the movement makes the task accessible across skill levels and reveals how submovements, such as rapid speed adjustments, drive mutual adaptation \cite{tomassini2022interpersonal}.

Using this setup, Alderisio et al. \cite{alderisio2017interaction} showed that synchronization depends on the homogeneity of participants' oscillation frequencies and on the structure of their interaction graph. Coordination is also robust to perturbations such as the loss of eye contact, with participants retaining the synchronized pace for about 7\,s \cite{bardy2020moving}. Moreover, spontaneous leadership can emerge during group coordination, making some participants more influential than others \cite{calabrese2021spontaneous}, and faster individuals typically slow down toward their partners' natural frequency to maintain synchrony \cite{calabrese2022modeling}.

Although vision is a natural channel for coordination, it has important limitations. First, visual feedback is intrinsically slow, introducing a sensorimotor delay of 100--300\,ms \cite{calabrese2022modeling}. In addition, changes in gaze direction and attentional focus can cause individuals to overlook communication partners or process visual information  intermittently or incompletely. Some task-relevant information may also be difficult or impossible to infer visually due the nature of the task, occlusions, or limited visibility. These limitations motivate the use of additional sensory channels able to transmit information about partners' movements more continuously and directly.

Haptic feedback is gaining attention as a means to communicate emotions and support naturalistic social interactions in both virtual reality \cite{jacucci2024haptics} and physical environments. Haptics offers a design advantage over visual or auditory feedback: the content, direction, and intensity of interaction can be explicitly controlled \cite{rognon2022linking}. Previous work shows that dyads of people spontaneously leverage haptic communication to simplify the task and reduce effort\cite{de2026haptic}, also sharing motion plans through interaction forces to improve joint tracking \cite{takagi2018haptic, Takagi2019}. Light tactile finger feedback can further enable synchronization in pairs of dancers \cite{sofianidis2014can}, facilitating motor coordination, increasing balance confidence, and ultimately improving postural stability and movement performance. Beyond dyads, haptic communication can induce teamwork in joint group activities such as team rowing \cite{tian2023group}. However, the role of haptic communication in inducing synchronization in human groups remains largely unexplored.

In this work, we investigate haptic communication as a means to enable synchronization in group motor tasks. We conducted an experiment with four people performing a joint oscillatory motion, rotating a robot handle continuously using wrist flexion--extension. The robots exerted torques on the handles as if they were connected by virtual elastic bands. To compare sensory modalities, participants performed the same task with haptic feedback, visual feedback, combined haptic-visual feedback, or no interaction. Visual feedback was provided by showing partners' motion on each participant's personal display. We tested communication in a ring topology, in which each participant was coupled only with their two neighbors, to assess how group coordination arises as information spreads over the network.

The contributions of this work are threefold. First, we introduce an experimental framework to investigate synchronization in groups interacting through haptic communication. Second, we assess how haptic feedback supports group synchronization, examining its potential as an alternative or complement to vision. Third, we compare haptic and visual communication modalities in group synchronization tasks, characterizing how sensory modalities affect the synchronization level, temporal stability, synchronization frequency, and movement smoothness. 

\section{Experimental design and Methods}
\label{sec:methods}

\subsection{Participants}
The study was approved by the ethics committee at Imperial College London (7112072). Twenty-four participants (11 female, 13 male), aged between 20 and 41 years, were recruited for this study. All participants were informed about the experiment, provided written informed consent, and completed a demographic questionnaire before participation. They were randomly assigned to six groups of four members each.
\begin{figure*}[t]
  \centering
    \includegraphics[width=\textwidth]{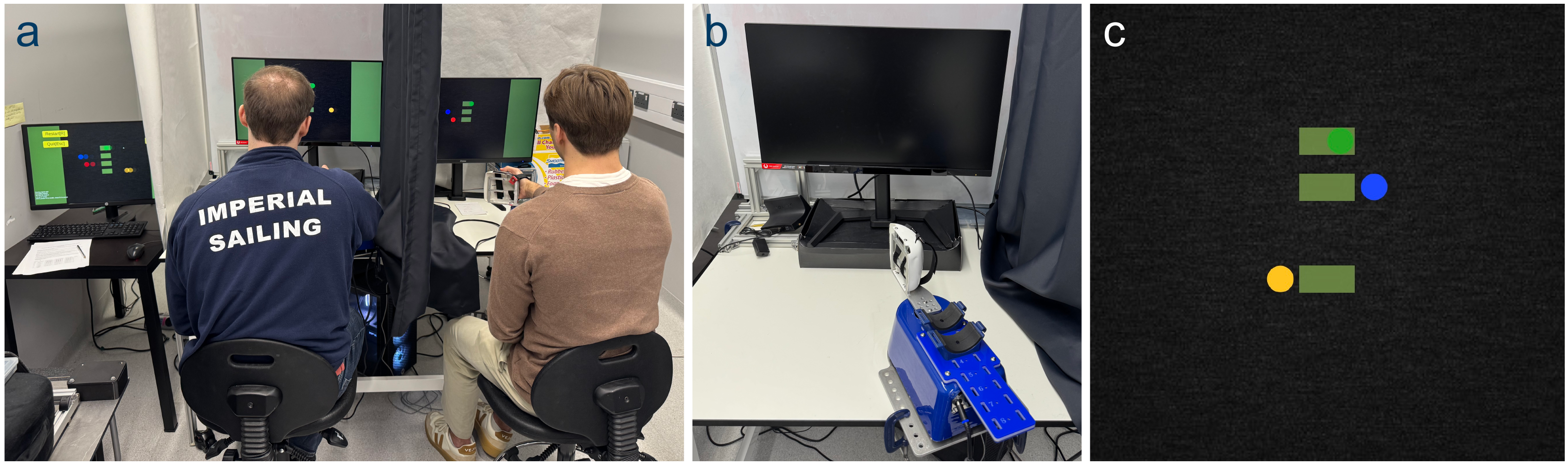}
  \caption{Experimental set-up. a: Two of the four participants performing the task, each using a personal HRX-1 robotic interface and an individual monitor as shown in panel b. Participants are separated by curtains to prevent viewing other participants' monitor and hand. The remaining two group members perform the task simultaneously beyond the dividers (not visible). The screen on the left of a is for the experimenter. c: Participant 1's display during visual interaction conditions (V and HV modes), showing their own cursor (green) and the cursors of participants 2 and 4 (blue and yellow, respectively), who are their neighbors in the ring topology. Participants move the cursor across the screen by manipulating the white handle via wrist flexion-extension movements. The central green rectangles indicate the starting area.}
  \label{fig:experimental_photos}
\end{figure*}

\subsection{Experimental setup} 

The experimental setup is shown in Figure~\ref{fig:experimental_photos}. We used four HRX-1 one-degree-of-freedom wrist robotic interfaces (HumanRobotiX, UK \cite{Farkhatdinov2026}) and four Iiyama G-Master GB2770HSU-B5 monitors (Fig.\,\ref{fig:experimental_photos}b). The robots were employed to provide interaction torque between participants and recorded wrist flexion/extension motion. 
Each robot was actuated by a brushless motor capable of exerting torques up to 4\,Nm,  was controlled by an Epos4 motor controller (Maxon Motors, Switzerland), and measured the wrist angle via an embedded optical encoder with a resolution 0.01$^\circ$. Each robot was fixed to the table at an angle of approximately $70^\circ$ relative to its front edge (Fig.\,\ref{fig:experimental_photos}b), affording the comfortable posture generally adopted for handwriting. Handle angular positions and torques were transmitted and recorded at 120\,Hz over a CAN bus, through a pCAN interface and a Texas Instruments C2000 microcontroller unit (MCU).

Monitors provided visual feedback on the flexion/extension motion. Each panel had a 27" diagonal, a resolution of 1920$\times$1080 pixels, and a maximum refresh rate of 165\,Hz. The custom visual interface was implemented in Unity (Figure~\ref{fig:experimental_photos}c), building on \emph{Chronos}, a platform for experiments on human group coordination in one-dimensional movements \cite{alderisio2017novel}.

During each trial, the four participants sat at their stations as shown in Figure~\ref{fig:experimental_photos}a, each equipped with a robot and a monitor. Curtains and separators prevented them from seeing the hands or screens of other group members. Each participant's wrist was secured to the ergonomic handle with straps. The handle was rotated through wrist flexion/extension and the resulting wrist motion was displayed on the screen as a horizontally moving cursor (Fig.\,\ref{fig:experimental_photos}c).

\subsection{Task and conditions} 
\label{sec:tasks_and_conditions}

Participants were instructed to continuously flex and extend their right wrist while coordinating with their partners' movements, following the feedback available in each experimental condition. Depending on the condition, they were asked to coordinate with the robot-generated torques, to synchronize with the additional cursors representing their partners' wrist positions, or with both. 
At the beginning of each trial, participants placed their wrist in a relaxed position, corresponding to the cursor being within the green rectangle in the middle of the screen (Fig.\,\ref{fig:experimental_photos}c). A 3\,s countdown was then displayed, after which they began moving, always starting with wrist flexion.

We tested four experimental feedback conditions: \emph{solo} (S), \emph{haptic} (H), \emph{visual} (V), \emph{haptic and visual} (HV):
\begin{itemize}
    \item [S:] Participants did not receive any feedback from the other quartet members, thus they saw only their own cursor and the robots exerted no (active) torque;
    \item [H:] Participants received torque feedback from their robot while seeing only their own cursor on the screen;
    \item [V:] Participants received no torque but saw their own cursor and the cursors of their two neighbors in real time;
    \item [HV:] Participants received both torque and visual feedback from neighbors.
\end{itemize}

\begin{figure}[t]
    \centering
    \includegraphics[width=0.75\columnwidth]{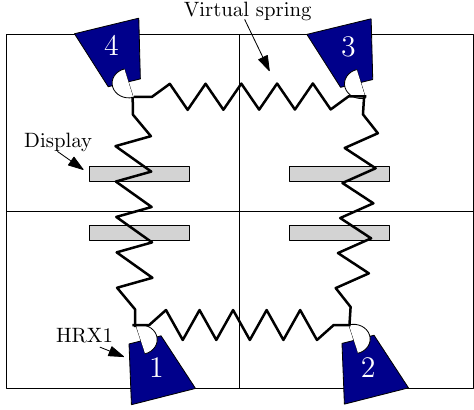}
    \caption{Schematic of the task setup. Each participant used a workstation equipped with an HRX-1 robot and a display. Participants controlled the robot by moving its handle using wrist flexion-extension movements. In H and HV modes (cf.~Section~\ref{sec:tasks_and_conditions}), the participant’s hand was coupled to their two neighbors via virtual elastic bands.}
    \label{fig:task_scheme}
\end{figure}

In conditions H and HV, letting $\phi_i$ denote the angular position of the $i$-th handle, the torque applied to the $i$-th participant was
\begin{equation}\label{eq:haptic_protocol}
    \tau_i = k \left[(\phi_{i+1}-\phi_i) + (\phi_{i-1}-\phi_i)\right],
\end{equation}
where $k>0$ is the stiffness coefficient and indices are taken modulo 4 over $\{1,2,3,4\}$, so that $4+1=1$ and $1-1=4$. Intuitively, each robot is pulled toward the angular positions of its two neighbors with a torque proportional to the position differences. This yields the ring interaction topology shown in Figure~\ref{fig:task_scheme}, in which each participant is coupled only to their two neighbors. In V and HV modes, each participant could see on their monitor both their own cursor and the cursors of their two neighbors in real time.

We tuned the virtual springs to deliver a soft coupling, perceptible to participants but not stiff enough to impose their movement. This choice reflects the study's goal of using haptics to stimulate group coordination rather than constrain individual motion. After preliminary tests, the stiffness coefficient in \eqref{eq:haptic_protocol} was set to $k = 0.1$\,Nm/rad, corresponding to a soft haptic coupling \cite{takagi2018haptic}.

Following prior studies on human group synchronization \cite{alderisio2017interaction, calabrese2021spontaneous, calabrese2022modeling}, participants performed 7 trials of 30\,s each for every condition, with a 10\,s rest interval between consecutive trials and conditions. Groups 1, 2, 3 performed the conditions in the order S--H--V--HV, while groups 4, 5, 6 executed them in the order S--V--H--HV.

\subsection{Phase of an oscillatory motion}
\label{subsec:freq_sync}

Let $\mathcal{T}\subseteq \mathbb{R}_{\geq 0}$ be a time interval and let $\mathbb{S} \coloneqq \mathbb{R}/2\pi\mathbb{Z}$ be the unit circle. The movement of the $i$-th participant is described by its \emph{oscillation phase} $\theta_i : \mathcal{T} \to \mathbb{S}$, which represents the progression of participant $i$ within the current wrist flexion-extension cycle, taken as one oscillation period. Namely, $\theta_i(t) = -\pi$ indicates the start of a period and corresponds to a (flexed wrist starting to extend); $\theta_i(t) = 0$ indicates the half of the period (wrist fully extended and starting to flex); $\theta_i(t) = \pi$ indicates the end of the period (wrist completes the flexion).

The phase of an oscillatory position signal can be computed offline through the Hilbert transform  \cite{pikovsky2001synchronization, king2009vol,spallone2026rope}. 
Given a signal $x:\mathcal{T}\to \mathbb{R}$ and its Hilbert transform $\mathcal{H}(x) : \mathcal{T}\to\mathbb{R}$, the \emph{analytic signal} at time $t \in \mathcal{T}$ is defined as $\xi(t)=x(t)+\mathrm{i}\mathcal{H}\left(x(t)\right)$.
The instantaneous phase $\theta : \mathcal{T} \to \mathbb{S}$ of $x$ is then
\begin{equation*}
    \theta(t) = \operatorname{arg}[\xi(t)] =\operatorname{atan2}\left[\mathcal{H} \!\left(x(t)\right), x(t)\right].
\end{equation*}
We applied this procedure to the recorded handle angular positions to obtain the instantaneous phase of each participant over time.

To compute the derivative of a phase signal $\theta_i$, we first \emph{unwrapped} it, removing the discontinuities occurring at the end of each cycle and yielding a continuous signal $\theta_i^{\mathrm{u}}:\mathcal{T} \to \mathbb{R}$.
The instantaneous frequency is then $\dot\theta_i = \mathrm{d}\theta_i^{\mathrm{u}} / \mathrm{d}t$. A group of $N$ participants is \emph{phase-synchronized} (exactly) at time $t$ if $\theta_i(t) = \theta_j(t), \forall i,j \in \{1,\ldots,N \}$, while they are \emph{frequency-synchronized} (exactly) if $\dot\theta_i(t) = \dot\theta_j(t), \forall i,j \in \{1,\ldots,N\}$. Derivatives were computed from phases data by forward finite differences and the resulting frequency signals were smoothed with a moving average filter over a sliding window of 10 samples with a stride of 1 sample, to attenuate measurement noise.

\begin{figure}[t]
    \centering
    \includegraphics[width= \columnwidth]{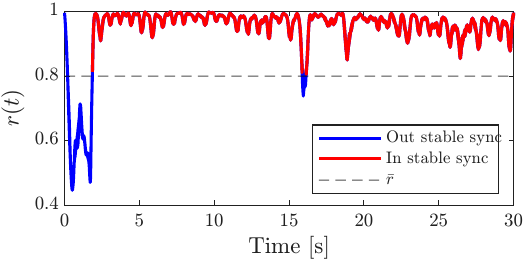}
    \caption{Evolution of the group order parameter magnitude $r(t)$ during one trial (group 3, mode H, trial 4). Red segments indicate stable synchronization windows $\Delta\mathcal{T}_i$ (see Section~\ref{sec:duration_sync}), defined as intervals where $r(t)\geq\bar{r}$ for at least $t_{\min} = 2$\,s. Blue segments denote the remaining periods, where $r(t) < \bar{r}$ or $r(t)\geq\bar{r}$ for less than $t_{\min} = 2$\,s. In this trial, $T_{\mathrm{sync}}= 0.93$.}
    \label{fig:time_sync_explained}
\end{figure}
\subsection{Synchronization metrics}
\label{subsec:metrics}

Across the experimental conditions, we compared the level of group synchronization and its temporal stability, quantified respectively by the \emph{order parameter} and the \emph{normalized duration of synchronization}.
\newline

\subsubsection{Order parameter}

The degree of synchronization was quantified using the \emph{Kuramoto order parameter}, a standard measure of phase alignment in networks of oscillators \cite{strogatz2000kuramoto, alderisio2017interaction}. 
For a group of $N \in \mathbb{N}$ oscillators with phases $\theta_j(t)$, it is defined as
\begin{equation}\label{eq:order_parameter}
    r(t) \, e^{\mathrm{i}\psi(t)} \coloneqq \frac{1}{N}\sum_{j=1}^{N}e^{\mathrm{i}\theta_j(t)},
\end{equation}
where $\psi(t)\in\mathbb{S}$ is the average phase at time $t$, and $r(t) \in [0, 1]$ measures phase synchronization, ranging from complete incoherence (e.g., phases uniformly distributed over the unit circle) at $r = 0$ to exact phase synchronization at $r = 1$. At frequency-synchronization, $\dot{r}(t)=0$.
We denote the time average of the order parameter magnitude in a trial by $\langle r \rangle$ and its standard deviation in time by $\sigma(r)$. 
\newline

\subsubsection{Normalized duration of synchronization}
\label{sec:duration_sync}

To quantify the stability of synchronization during a trial of duration $T > 0$, we introduce the \emph{normalized duration of synchronization} $T_{\mathrm{sync}} \in [0,1]$, defined as the fraction of the trial in which the order parameter magnitude remained above a threshold $\bar{r} \in (0, 1)$ for at least a minimum time $t_{\min} \in (0, T)$ so that brief intervals of accidental alignment are excluded. Formally, let $\Delta \mathcal{T}_i$ denote the maximal time intervals over which $r(t)\geq \bar{r}$. Each interval starts when $r(t)$ crosses $\bar{r}$ from below and ends when it next drops below $\bar{r}$; consecutive intervals are therefore disjoint and separated by intervals in which $r(t) < \bar{r}$. Letting $|\cdot|$ denote the measure of a set, we define $\mathcal{T}_\mathrm{sync} \coloneqq \bigcup_{i \,:\, |\Delta \mathcal{T}_i| \geq t_{\min}} \Delta \mathcal{T}_i$ and $T_{\mathrm{sync}} \coloneqq \frac{1}{T} \left\lvert \mathcal{T}_\mathrm{sync} \right\rvert$. $T_{\mathrm{sync}} = 1$ means persistent synchronization ($r(t) \ge \bar{r}, \forall t$), while $T_{\mathrm{sync}} = 0$ indicates absence of meaningful synchronization ($r(t)$ never exceeds $\bar{r}$ for at least $t_{\mathrm{min}}$). 

In this study, $T=30$\,s. Moreover, we set $\bar{r} = 0.8$ following \cite{bardy2020moving, shahal2020synchronization}, where this value is used to mark \emph{weak synchronization}. The value of $t_{\mathrm{min}}$ was set empirically to 2\,s after inspecting the order parameter magnitude $r(t)$ across trials, groups, and feedback conditions. Brief excursions above $\bar r$ typically lasted less than 2\,s and were not followed by sustained alignment, whereas genuine coordination, once established, persisted well beyond this interval. This choice therefore excludes transient threshold crossings from the estimate of sustained synchronization. Figure~\ref{fig:time_sync_explained} illustrates the computation of $T_{\mathrm{sync}}$ on a representative trial.

\subsection{Motion smoothness}
\label{sec:smoothness}

A repetitive motion is called \emph{smooth} if, in loose terms, it is continuous and non-intermittent, independently of its amplitude and duration \cite{balasubramanian2015analysis}; reduced smoothness is associated with poorer control ability  \cite{balasubramanian2015analysis}. In wrist flexion/extension, intermittency appears as repeated alternation between acceleration and deceleration within a half-cycle, or as finite intervals of null velocity.

Motion smoothness was measured using the \emph{spectral arc length} (SPARC) \cite{balasubramanian2015analysis}, the arc length of the normalized Fourier magnitude spectrum of a velocity signal. Formally, let $v:\mathcal{T} \to \mathbb{R}$ be a velocity signal, $V(\omega)$ denotes its Fourier magnitude spectrum and $\hat{V}(\omega)\coloneqq {V(\omega)}/{V(0)}$. Given thresholds $\bar{V}$ and $\omega_\mathrm{c}^{\max}$, define the cutoff frequency 
\begin{equation}
    \omega_{\mathrm{c}} \coloneqq \min \{\omega_\mathrm{c}^{\max}, \min\{\omega \mid \forall \nu > \omega, \hat{V}(\nu) < \bar{V}\}\}.
\end{equation}
and compute the SPARC $S(v) < 0$ as
\begin{equation}
    \label{eq:sparc}
    S(v) \coloneqq -\int_{0}^{\omega_{\mathrm{c}}} \! \left[\left(\frac{1}{\omega_\mathrm{c}}\right)^{\!\!2}+\left(\frac{d\hat{V}(\omega)}{d\omega}\right)^{\!\!2} \, \right]^{\frac{1}{2}} \!\!\! \mathrm{d}\omega \,.
\end{equation}
In simple terms, a movement whose velocity spectrum is concentrated in few frequencies results in a shorter spectral arc length and a SPARC value that is smaller in magnitude, whereas movement intermittencies introduce additional spectral components and fluctuations, lengthen the arc, and drive $S(v)$ more negative.

We performed smoothness analysis on angular velocities signals, obtained offline from the recorded handle angular positions by forward finite differences followed by a moving average filter over a sliding window of 10 samples with a stride of 1 sample. Following \cite{balasubramanian2015analysis}, we segmented the trial velocity signal $v(t)$ into its periods $v_m$, $m \in \{1, \dots, n_\mathrm{p}\}$, detecting a new period at each phase transition from $\pi$ to $-\pi$. We then computed $S(v_m)$ for each period as in \eqref{eq:sparc} and averaged across periods, obtaining
$S(v) = \frac{1}{n_\mathrm{p}} \sum_{m=1}^{n_\mathrm{p}} S(v_m)$.
We used the code provided in \cite{balasubramanian2015analysis}, with parameters $\omega_\mathrm{c}^{\max} = 20$~Hz and $\bar{V} = 0.05$ as recommended therein.

\begin{figure}[t]
  \centering
    \includegraphics[width=0.9\columnwidth]{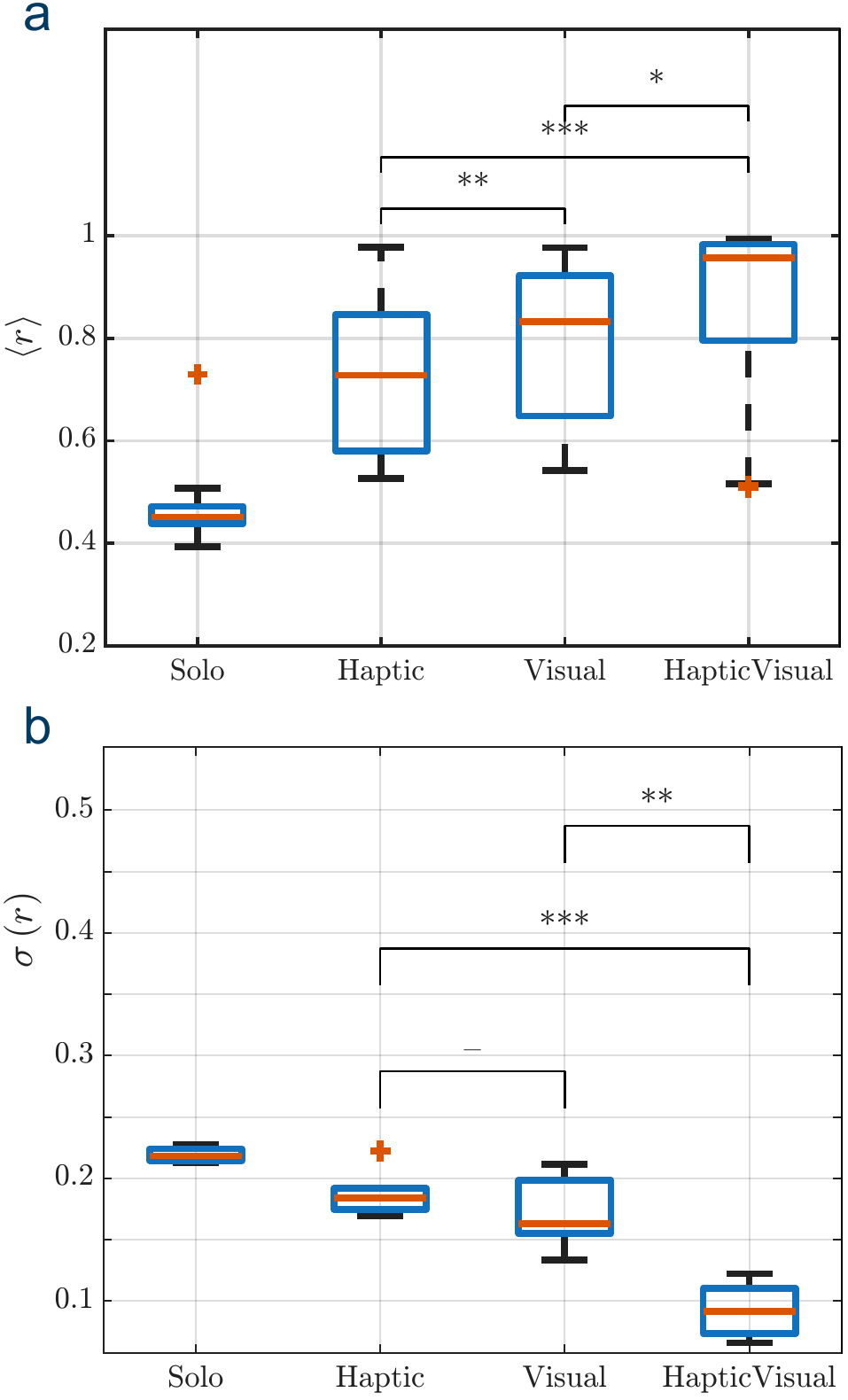}
    \caption{Comparison of the time-averaged order parameter magnitude across modes (a) and its standard deviation (b). Boxplots show the distribution across all trials from all groups within each mode. The significance of one tailed t-tests between corresponding modes is indicated as -- for $p \geq 0.05$, * for $p<0.05$, ** for $p<0.01$ and *** for $p<0.001$.}
  \label{fig:order_param_analysis}
\end{figure}

\subsection{Statistical comparisons}

The four experimental conditions (cf.~Sections~\ref{sec:tasks_and_conditions}) were compared using the metrics described in Sections~\ref{subsec:metrics} and~\ref{sec:smoothness}. Pairwise differences between conditions were assessed with one-tailed $t$-tests. To control the false discovery rate across multiple comparisons, we applied the Benjamini-Hochberg procedure  \cite{benjamini1995controlling}.

\section{Results}

\subsection{Effects of sensory feedback on synchronization level}
Figure\,\ref{fig:order_param_analysis}a shows the distributions of the time averaged order parameter magnitude $\langle r \rangle$  across the four conditions of Section~\ref{sec:tasks_and_conditions}, pooled over all groups and trials. All feedback conditions (H, V, HV) yielded markedly higher synchronization than condition~S ($p<0.001$ for all comparisons). HV achieved the highest synchronization level, with significantly larger $\langle r \rangle$ than both H ($p = 0.036$) and V ($p = 0.032$). Visual feedback (V) also led to higher synchronization than haptic feedback alone (H) ($p = 0.0006$).

\begin{figure*}[t]
    \centering
    \includegraphics[width=0.95\textwidth]{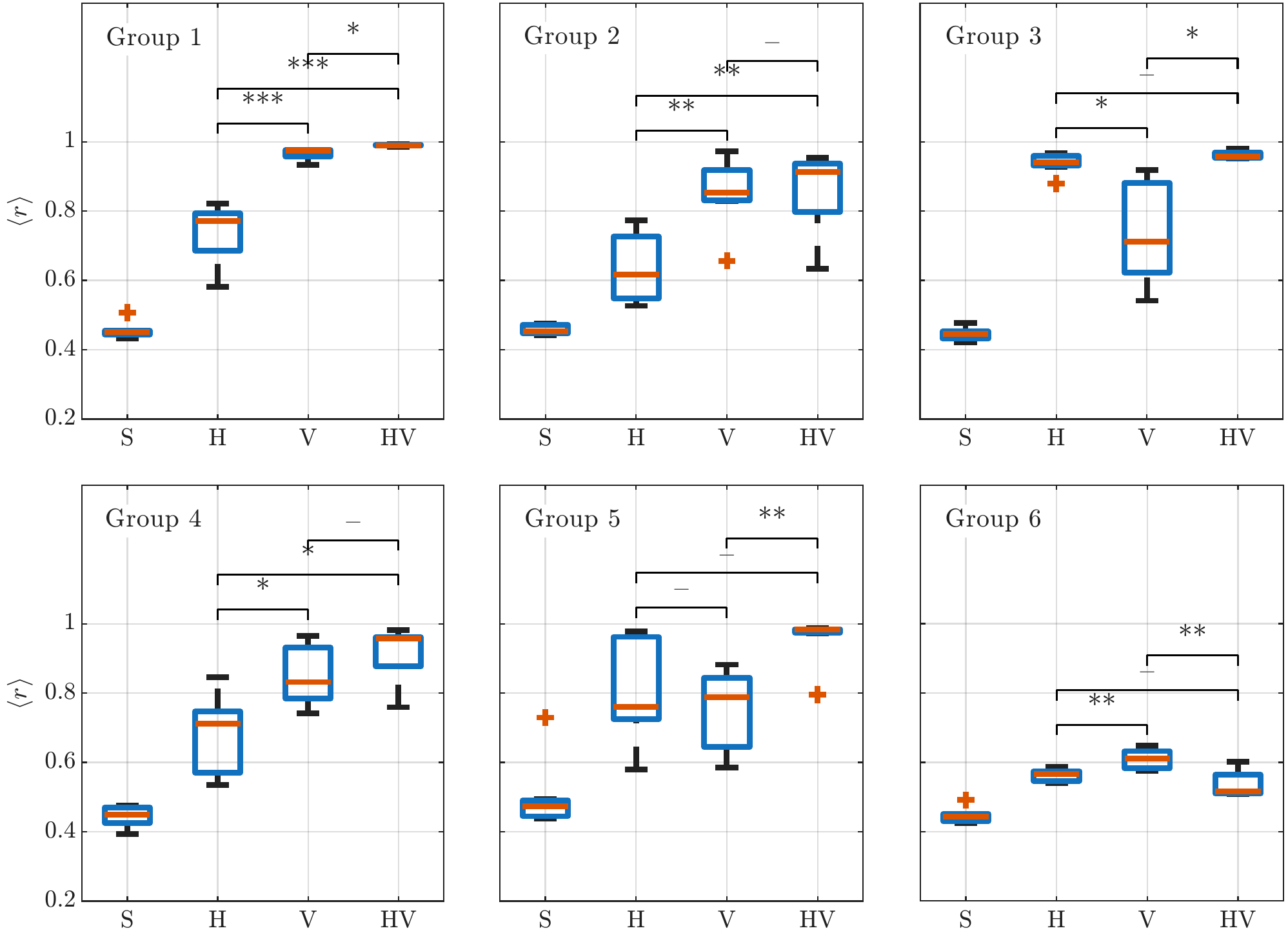}
    \caption{Comparison of time-averaged order parameter magnitude across modes for each group. Boxplots show the distribution over all trials performed by that group in each mode. Modes: S = solo, H = haptic, V = visual and HV = haptic \& visual. The significance of one tailed t-tests between corresponding modes is indicated as -- for $p \geq 0.05$, * for $p<0.05$, ** for $p<0.01$ and *** for $p<0.001$.}
    \label{fig:order_parameter_mean_groups}
\end{figure*}

The corresponding distributions of $\sigma(r)$, the temporal standard deviation of $r(t)$, are shown in Figure\,\ref{fig:order_param_analysis}b. HV produced the lowest temporal variability, lower than both H ($p = 0.0003$) and V ($p = 0.004$), whereas H and V did not differ significantly ($p > 0.05$).

Figure\,\ref{fig:order_parameter_mean_groups} shows $\langle r \rangle$ for each group. Group 6 appears as an outlier, with minimal synchronization across conditions. Among the remaining groups, HV either achieved the highest synchronization or matched the best-performing condition.

\begin{figure}[t]
    \centering
    \includegraphics[width=0.9\linewidth]{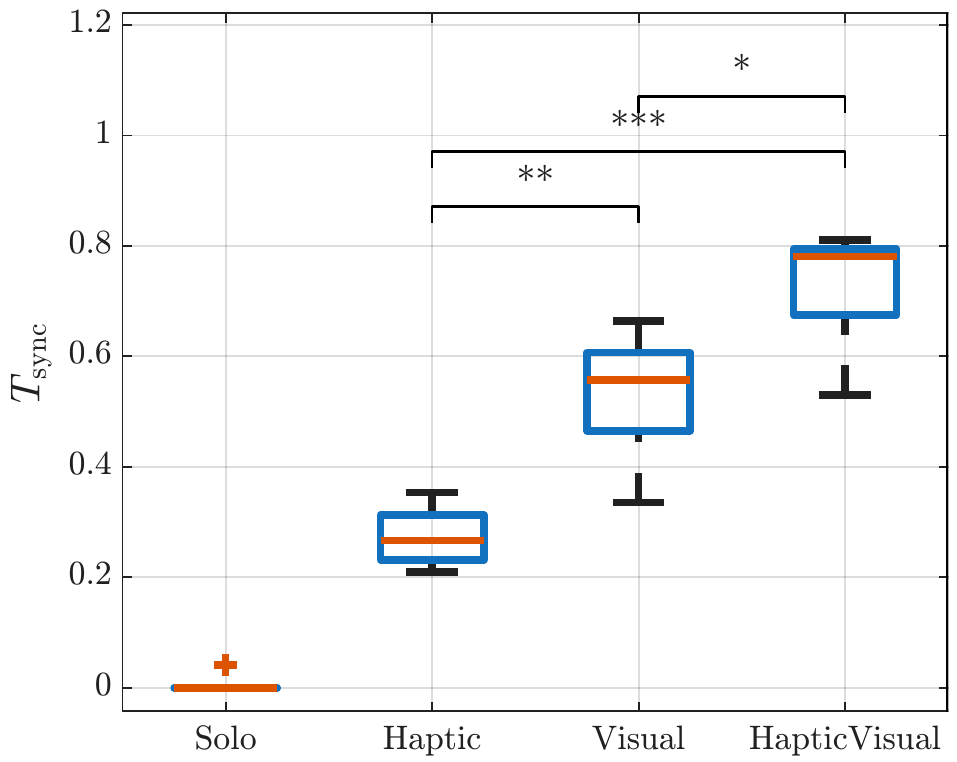}
\caption{Comparison of synchronization duration $T_{\mathrm{sync}}$ across modes. Boxplots show the distribution across all trials from all groups within each mode. The significance of one tailed t-tests between corresponding modes is indicated as -- for $p \geq 0.05$, * for $p<0.05$, ** for $p<0.01$ and *** for $p<0.001$. Tests with the solo condition are omitted for clarity.}
    \label{fig:duration_of_sync}
\end{figure}

\subsection{Effects of sensory feedback on synchronization duration}

Figure~\ref{fig:duration_of_sync} compares the distributions of the synchronization duration $T_{\mathrm{sync}}$ across conditions. HV trials showed longer synchronization intervals than both H ($p = 0.0002$) and V ($p = 0.032$), with a median of approximately $0.8$ (about 24\,s of sustained synchronization within the 30\,s trials). Visual feedback (V) also resulted in longer synchronization durations than haptic feedback (H) alone ($p = 0.0011$). Disaggregating by group yields a pattern qualitatively similar to Figure\,\ref{fig:order_parameter_mean_groups}: HV outperformed H in groups~1, 2, 4 and outperformed V in groups~3, 5, while group~6 briefly synchronized only in one trial in condition~V. The corresponding figure is omitted for brevity.

\begin{figure*}[!t]
    \centering
    \includegraphics[width=\textwidth]{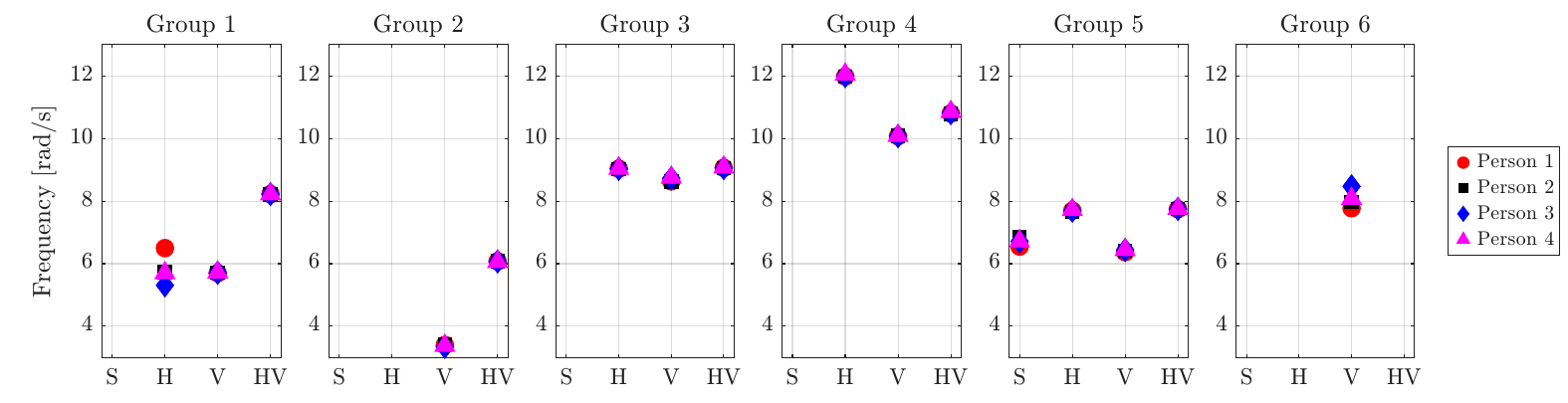}
    \caption{Comparison of participants' synchronization frequency across modes, shown separately for each group. Each point represents a participant's average synchronization frequency computed over the time windows included in $\mathcal{T}_{\mathrm{sync}}$. No marker is shown for modes and groups in which high synchronization was not stably reached (i.e., $T_{\mathrm{sync}} = 0$). Modes: S = solo, H = haptic, V = visual, HV = haptic \& visual.}
    \label{fig:frequency_comparison}
\end{figure*}
\subsection{Effect of haptic feedback on synchronization frequencies}
\label{sec:effect_haptics_on_frequency}

For each participant, let $\langle \dot\theta_i \rangle_{\mathcal{T}_{\mathrm{sync}}}$ denote the average instantaneous frequency $\dot\theta_i$ over $\mathcal{T}_\mathrm{sync}$ (cf.~Section~\ref{sec:duration_sync}). Figure~\ref{fig:frequency_comparison} reports this quantity, averaged over trials of each condition, for every participant and group. For each condition $m \in \{\mathrm{S}, \mathrm{H}, \mathrm{V}, \mathrm{HV}\}$ and group $g \in \{1,\dots,6\}$, we define the \emph{average synchronization frequency} $\bar{\omega}_g^m$ as the average of $\langle \dot\theta_i \rangle_{\mathcal{T}_{\mathrm{sync}}}$ over the $N=4$ group members. We then computed the group-wise percentage increase between two conditions $m, m' \in \{\mathrm{H},\mathrm{V},\mathrm{HV}\}$ as
\begin{equation*}
    I_g^{(m,m')} \!= 
    \frac{\bar{\omega}_g^{m}-\bar{\omega}_g^{m'}}
    {\bar{\omega}_g^{m'}} 100 \,,
\end{equation*}
and averaged it over the groups that reached synchronization in both conditions (i.e., excluding cases with $\mathcal{T}_\mathrm{sync} = \varnothing$ in either one), denoting the result by $\langle I^{(m,m')} \rangle_g$. Synchronization frequencies were higher whenever haptic feedback was present: $\langle I^{(\mathrm{H},\mathrm{V})} \rangle_g = 11.2\%$ (H higher than V), $\langle I^{(\mathrm{HV},\mathrm{H})} \rangle_g = 8.2\%$ (HV higher than H), and $\langle I^{(\mathrm{HV},\mathrm{V})} \rangle_g = 31.2\%$ (HV higher than V).

\subsection{Motion smoothness depends on movement frequency}

Motion smoothness, quantified via the SPARC metric (Section \ref{sec:smoothness}) is shown in Figure~\ref{fig:smoothness_comparison}. Fig.\,\ref{fig:smoothness_comparison}a shows that HV trials were smoother than both H ($p = 0.0002$) and V ($p = 0.0045$), while H and V did not differ significantly. Figure~\ref{fig:smoothness_comparison}b pools all conditions and shows that smoothness increases sharply with the trial's mean movement frequency and levels off. An exponential least-squares fit computed as in \cite{coleman1996interior} yields the SPARC values
\begin{equation*}
    S_i=-4.0212e^{-0.835\langle\dot\theta_i\rangle}-2.5829,
\end{equation*}
with a residual norm of 33.5895. 

\section{Discussion}

\subsection{Haptic feedback enables group synchronization}

A first finding of this study is that haptic interaction via virtual elastic bands can induce frequency synchronization among people performing repetitive motion, even without visual cue. As shown in Figure~\ref{fig:order_param_analysis}a, $\langle r \rangle$ is consistently higher in condition~H than in condition~S, in which participants received no feedback on their partners' motion. Because the selected coupling stiffness of $k = 0.1$\,Nm/rad was well below passive wrist stiffness (flexion: 0.55\,Nm/rad; extension:  1.02\,Nm/rad  \cite{formica2012passive}), the virtual springs could perturb but not drive the limb. Therefore, the observed synchronization is unlikely to be passive mechanical entrainment and instead indicates voluntary alignment to the torque cues.

\begin{figure}[!t]
  \centering
    \includegraphics[width=0.9\columnwidth]{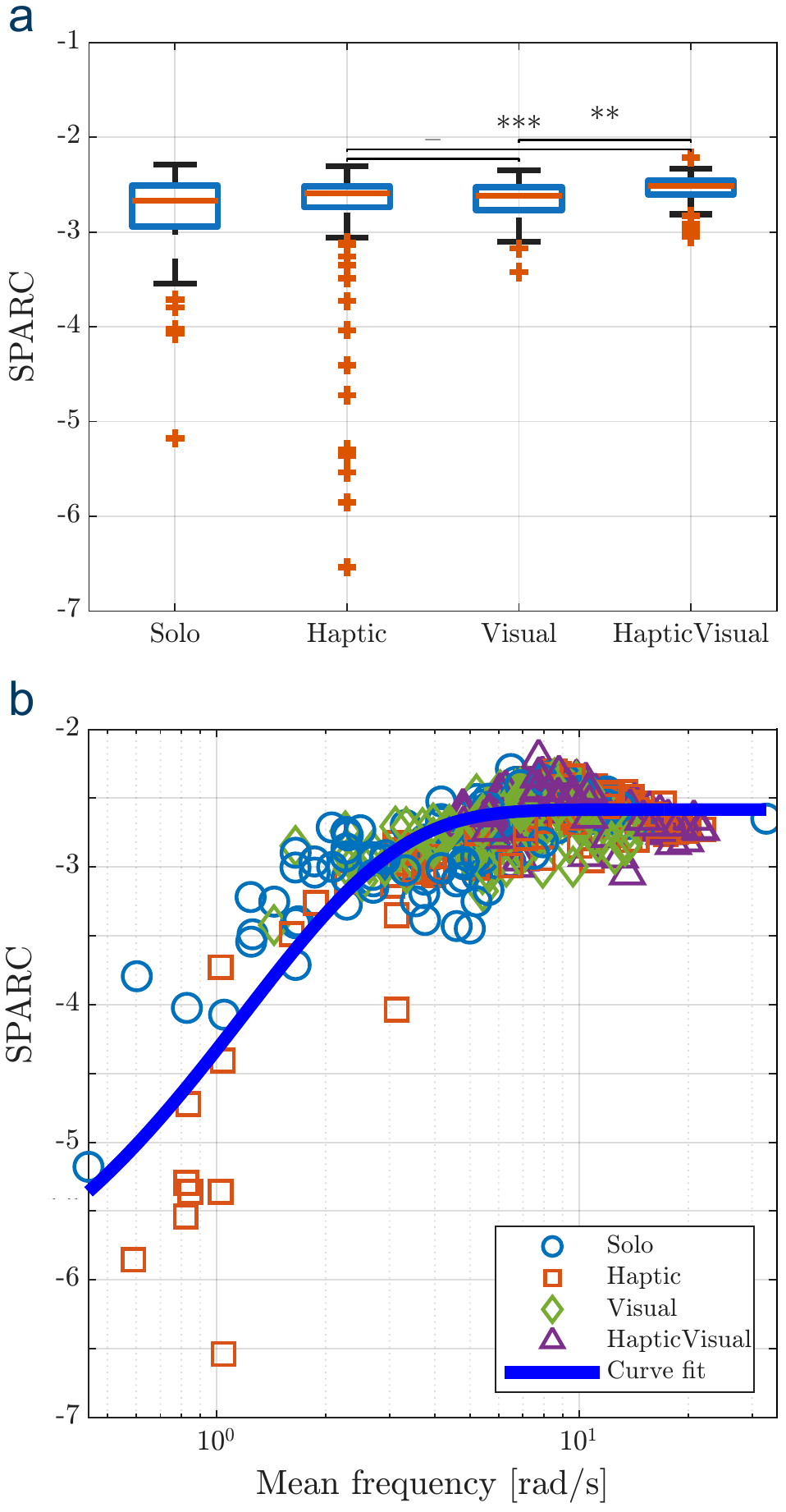}
    \caption{Motion smoothness analysis. (a) Comparison of SPARC metrics across modes. The significance of one tailed t-tests between corresponding modes is indicated as -- for $p \geq 0.05$, * for $p<0.05$, ** for $p<0.01$ and *** for $p<0.001$; tests involving the solo condition are omitted for clarity. (b) Trial smoothness as a function of average frequency. Each point represents a single trial from any group and mode; the blue line is an exponential fit to the data.}
  \label{fig:smoothness_comparison}
\end{figure}

\subsection{Comparison between haptic and visual feedback}

Although haptic feedback was sufficient to induce synchronization, visual interaction yielded, on average, higher and more stable synchronization (Figs.\,\ref{fig:order_param_analysis}a,\ref{fig:duration_of_sync}), suggesting that participants could more readily exploit visual information for coordination in the present task. However, disaggregating by group nuances this picture (Fig.\,\ref{fig:order_parameter_mean_groups}). Group~6 almost never reached sustained synchronization, in any condition; which may reflect poor understanding of the instructions or a lack of engagement. Groups {1,2,4} synchronized better under visual than haptic feedback, group~3 showed the converse, and group~5 showed no significant difference between the two. This split does not align with the order in which conditions were administered (cf.~Section~\ref{sec:tasks_and_conditions}): groups~1 and~2 received H before V while group~4 received V before H; likewise for groups~3 and~5. Some groups therefore achieved high synchronization through weak haptic coupling alone, making haptic feedback a viable alternative where the visual field is obstructed or unavailable.

\subsection{Combining haptic and visual feedback improves precision and stability}

The strongest synchronization performance was obtained when both haptic and visual feedback were provided. Overall, HV trials outperformed both H and V in degree of synchronization ($\langle r \rangle$), temporal variability ($\sigma(r)$), and synchronization duration ($T_{\mathrm{sync}}$), with all comparisons significant (Figures~\ref{fig:order_param_analysis} and~\ref{fig:duration_of_sync}). At the group level, HV either achieved the highest $\langle r \rangle$ or matched the best-performing condition in every group (Fig.\,\ref{fig:order_parameter_mean_groups}), except group~6, which seldom synchronized in any condition.

A plausible explanation for these results is that the two channels carry complementary information. Vision provides an explicit representation of partners' motion, allowing participants to assess the overall state of group coordination. Haptics provides continuous force cues proportional to relative position differences, supporting finer adjustments of the ongoing movement. Combining them grants access to both, accounting for the HV synchronization gains over either modality alone.

\subsection{Haptic feedback affects synchronization frequency}

Prior work on visually mediated coordination has shown that groups tend to settle at frequencies slower than the natural pace of their faster members \cite{bardy2020moving, calabrese2022modeling}. Our results add that the settled frequency also depends on the interaction modality, being consistently higher when haptic feedback is available than under visual feedback alone. One possible explanation is that mechanical coupling engages fast proprioceptive feedback loops: muscle responses to limb perturbation appear at short latency (20--50\,ms) and long latency (50--100\,ms) \cite{pruszynski2012optimal}, whereas responses to visual perturbations of the hand appear in muscle activity only after about 110\,ms and in force at the hand after about 150\,ms \cite{franklin2008specificity}. Participants coupled through the virtual springs could thus compensate relative motion at oscillation frequencies where visual feedback arrives too late to be effective.

\subsection{Movement smoothness is linked to movement frequency}

A further finding is that the smoothness of oscillatory movement covaries with the movement frequency. As shown in Figure~\ref{fig:smoothness_comparison}b, smoothness increases sharply with a trial's mean frequency and levels off beyond a certain value, following an exponential trend. This saturation of the smoothness metric is due to the task requiring intermittent control to reverse direction once per half-period.

\section{Conclusion}

We investigated how haptic interaction affects synchronization in groups performing repetitive motion, a setting in which visual interaction has been well studied, while haptics has been explored mostly in dyads. Six quartets attempted to synchronize during continuous oscillatory wrist flexion-extension under haptic feedback (through virtual elastic bands), visual feedback, both, or neither.

We found that weak haptic coupling to only the two neighbors was sufficient to induce group synchronization, without visual cues. Visual interaction yielded higher and more stable synchronization than haptics alone on aggregate, although the two modalities were comparable in some groups. Combining them outperformed either one alone in synchronization level, temporal stability, and duration, indicating that the two channels carry complementary information. Haptic feedback also influenced the movement itself, raising the frequency at which groups coordinated. Finally, movement smoothness covaried with that frequency.

The scope of these results is limited by two design aspects. First, the stiffness of the virtual springs was fixed at a single low value. Second, participants were explicitly instructed to synchronize, isolating the effect of sensory modalities but making success depend on understanding and compliance, as illustrated by group~6, which rarely reached sustained synchronization. These limitations suggest directions for future work. Varying stiffness, communication delay, and interaction topology would clarify how haptic-coupling parameters shape group behavior. Moreover, haptic feedback affords unique design opportunities: unlike visual feedback, haptic torques reach the limb regardless of attentional focus, allowing control over what is transmitted, that it is received, and with what intensity and timing. This enables real-time protocols that adapt feedback to each group member---potentially via virtual agents embedded in the group  \cite{grotta2024learning, diporzio2025personalized}---to enhance synchronization. Finally, to assess ecological validity, these effects should be tested in tasks where synchronization serves an external goal (e.g., team rowing, collaborative manipulation, or group rehabilitation), rather than being the goal itself.

\bibliographystyle{IEEEtran} 
\bibliography{IEEEabrv,bibliography}
\vfill

\end{document}